\documentclass[amsmath,aps,showpacs,a4paper,10pt]{revtex4}
 \usepackage{epsf}
 \usepackage{graphicx}    % Include figure files

\begin{document}

 \newcommand{\be}[1]{\begin{equation}\label{#1}}
 \newcommand{\ee}{\end{equation}}
 \newcommand{\bea}{\begin{eqnarray}}
 \newcommand{\eea}{\end{eqnarray}}
 \def\disp{\displaystyle}

 \def\gsim{ \lower .75ex \hbox{$\sim$} \llap{\raise .27ex \hbox{$>$}} }
 \def\lsim{ \lower .75ex \hbox{$\sim$} \llap{\raise .27ex \hbox{$<$}} }

 \begin{titlepage}

 \begin{flushright}
 arXiv:1207.2898
 \end{flushright}

 \title{\Large \bf Quasi-Rip: A New Type of
 Rip Model without~Cosmic~Doomsday}

 \author{Hao~Wei\,}
 \email[\,email address:\ ]{haowei@bit.edu.cn}
 \affiliation{School of Physics, Beijing Institute
 of Technology, Beijing 100081, China}

 \author{Long-Fei~Wang\,}
 \affiliation{School of Physics, Beijing Institute
 of Technology, Beijing 100081, China}

 \author{Xiao-Jiao~Guo}
 \affiliation{School of Physics, Beijing Institute
 of Technology, Beijing 100081, China}

 \begin{abstract}\vspace{1cm}
 \centerline{\bf ABSTRACT}\vspace{2mm}
 The fate of our universe is an unceasing topic of cosmology
 and the human being. The discovery of the current accelerated
 expansion of the universe significantly changed our view of
 the fate of the universe. Recently, some interesting scenarios
 concerning the fate of the universe attracted much attention
 in the community, namely the so-called ``Little Rip'' and
 ``Pseudo-Rip''. It is worth noting that all the Big Rip,
 Little Rip and Pseudo-Rip arise from the assumption that the
 dark energy density $\rho(a)$ is monotonically increasing. In
 the present work, we are interested to investigate what will
 happen if this assumption is broken, and then propose a
 so-called ``Quasi-Rip'' scenario, which is driven by a type
 of quintom dark energy. In this work, we consider an explicit
 model of Quasi-Rip in detail. We show that Quasi-Rip has an
 unique feature different from Big Rip, Little Rip and
 Pseudo-Rip. Our universe has a chance to be rebuilt from the
 ashes after the terrible rip. This might be the last hope in
 the ``hopeless'' rip.
 \end{abstract}

 \pacs{95.36.+x, 98.80.-k}

 \maketitle

 \end{titlepage}

 \renewcommand{\baselinestretch}{1.1}

%============================= section 1 ===================================

\section{Introduction}\label{sec1}

Since its discovery in 1998, the current accelerated expansion
 of our universe~\cite{r1} has been one of the most active
 fields in modern cosmology. As is well known, it could be due
 to an unknown energy component (dark energy) or a modification
 to general relativity (modified gravity)~\cite{r1,r2}. Before
 1998, it was commonly believed that in the future our universe
 will either expand forever or contract again into a final Big
 Crunch. However, the discovery of the current accelerated
 expansion of the universe significantly changed our view of
 the fate of the universe. In fact, many novel possibilities
 are under the active consideration in the community nowadays.

Today, there are many dark energy candidates in the market.
 Among them, the famous phantom dark energy~\cite{r3,r4} is
 very interesting. Its equation-of-state parameter (EoS) is
 smaller than $-1$. Although phantom dark energy is consistent
 with the current observational data~\cite{r1,r3}, it violates
 all the energy conditions. One of the consequences is that
 our universe will encounter a singularity at a finite time,
 namely the so-called Big Rip~\cite{r4}. At this singularity,
 the scale factor $a$, energy density and pressure of our
 universe are all divergent. In fact, besides the traditional
 Big Bang, Big Crunch, and the Big Rip, many novel
 singularities have been considered in the literature, such as
 Sudden singularities, Generalized sudden singularities,
 Quiescent singularities, Big Boost, Big Brake, Big Freeze,
 $w$ singularities, Inaccessible singularities, Directional
 singularities (see e.g.~\cite{r5,r6} and references therein
 for some brief reviews). These singularities arise at the
 price of violation of one or several energy conditions.
 As is well known, in~\cite{r6} (see also e.g.~\cite{r5,r7,r8})
 the future singularities have been classified into four types,
 namely
 \begin{itemize}
   \item Type I (Big Rip): $a\to\infty$, $\rho\to\infty$,
    $H\to\infty$, $|p|\to\infty$, when $t\to t_s<\infty\,$;
   \item Type II (Sudden singularity): $a\to a_s$, $\rho\to\rho_s$,
    $H\to H_s$, $|p|\to\infty$, $\dot{H}\to\infty$, when
     $t\to t_s<\infty\,$;
   \item Type III (Big Freeze): $a\to a_s$, $\rho\to\infty$,
    $H\to\infty$, $|p|\to\infty$, when $t\to t_s<\infty\,$;
   \item Type IV (Generalized sudden singularity): $a\to a_s$,
    $\rho\to\rho_s$, $H\to H_s$, $|p|\to p_s$,
     $\dot{H}\to\dot{H}_s$, and higher derivatives
     of $H$ diverge, when $t\to t_s<\infty\,$,
 \end{itemize}
 where $t_s$, $a_s$, $\rho_s$, $H_s$, $p_s$, $\dot{H}_s$ are
 all finite constants ($a_s\not=0$); $\rho$ and $p$ are energy
 density and pressure respectively; $H\equiv\dot{a}/a$ is the
 Hubble parameter, and a dot denotes a derivative with respect
 to the cosmic time $t$. In fact, the above four types can
 include almost all known future singularities.

Of course, singularities usually are not desirable in physics.
 Therefore, other possible fates of our universe are also
 considered in the literature, such as the cyclic/oscillatory
 cosmology. Recently, some interesting scenarios concerning the
 fate of the universe attracted much attention in the
 community, namely the so-called ``Little Rip''~\cite{r9} and
 ``Pseudo-Rip''~\cite{r10}. In~\cite{r9,r10,r11}, their authors
 showed that if the cosmic energy density will remain constant
 or monotonically increase in the future, then all the possible
 fates of our universe can be divided into four categories
 based on the time asymptotics of the Hubble parameter
 $H(t)$~\cite{r10}, namely
 \begin{itemize}
   \item Big Rip: $H(t)\to\infty$, when $t\to t_{rip}<\infty\,$;
   \item Little Rip: $H(t)\to\infty$, when $t\to\infty\,$;
   \item Cosmological Constant: $H(t)=const.\,$;
   \item Pseudo-Rip: $H(t)\to H_\infty <\infty$, when $t\to\infty\,$,
 \end{itemize}
 where $H_\infty$ is a constant. Obviously, the Big Rip singularity
 is not the only fate of our universe with the phantom-like
 dark energy. Both Little Rip and Pseudo-Rip are non-singular,
 and hence fall outside of the four categories in~\cite{r6}.
 Similar to the Big Rip, the Little Rip dissociates {\em all}
 bound structures, but the strength of dark energy is not
 enough to rip apart spacetime (unlike the Big Rip)~\cite{r9,r10}.
 On the other hand, the Pseudo-Rip dissociates the bound structures
 which are held together by a binding force at or below a
 particular threshold, and hence it is possible that only
 {\em some} bound structures are dissociated while the others
 are {\em not} dissociated (depending on the model
 parameters)~\cite{r10}. In fact, the Little Rip is an intermediate
 case between the cosmological constant and the Big Rip~\cite{r9},
 while the Pseudo-Rip is an intermediate case between the
 cosmological constant and the Little Rip~\cite{r10}.

It is worth noting that all the Big Rip, Little Rip and Pseudo-Rip
 arise from the assumption that the dark energy density
 $\rho(a)$ is monotonically increasing~\cite{r9,r10,r11}, i.e., the
 dark energy is phantom-like (its EoS $w<-1$). In the present work,
 we are interested to investigate what will happen if this
 assumption is broken. Obviously, in the case of the dark
 energy density $\rho(a)$ is monotonically decreasing (i.e.,
 the dark energy is quintessence-like with an EoS $w>-1$), no
 rip will happen and no bound structures will be dissociated.
 On the other hand, in the case of the dark energy density $\rho(a)$
 monotonically decreases (namely $w>-1$) in the first stage
 and then monotonically increases (namely $w<-1$) in the second
 stage (this is the case of the so-called ``quintom A'' dark energy
 in terminology of e.g.~\cite{r12} and references therein), the fate
 of our universe is the Big Rip, which is trivial in some sense. The
 third case is that the dark energy density $\rho(a)$ monotonically
 increases (namely $w<-1$) in the first stage and then
 monotonically decreases (namely $w>-1$) in the second stage
 (this is the case of the so-called ``quintom~B'' dark
 energy in terminology of e.g.~\cite{r12} and references
 therein). It can be expected that in the first stage some or
 all bound structures will be dissociated (similar to the case
 of Pseudo-Rip), but then the disintegration process will stop, and
 the already disintegrated structures have the possibility to
 be recombined in the second stage. We dub it ``Quasi-Rip'',
 which is the subject of the present work (here we temporarily
 do not consider the case of oscillatory quintom dark energy).
 Since the quintom-like dark energy~\cite{r13} (whose EoS can
 cross the so-called phantom divide $w=-1$) is slightly favored
 by the observational data~\cite{r1} (see e.g.~\cite{r12} for a
 comprehensive review), we note that the Quasi-Rip is well-motivated
 in fact.

This paper is organized as follow. In Sec.~\ref{sec2}, we
 discuss the disintegration of bound structures. In Sec.~\ref{sec3},
 we present an explicit model of Quasi-Rip. We constrain this model
 with the observational data, and then clearly show the Quasi-Rip in
 this model. In Sec.~\ref{sec4}, some concluding remarks are given.

%============================= section 2 ===================================

\section{The disintegration of bound structures}\label{sec2}

It is useful to introduce the concept of ``inertial force''
 when we discuss the disintegration of bound structures.
 According to~\cite{r10,r11,r23}, in a flat
 Friedmann-Robertson-Walker (FRW) universe dominated by dark energy,
 (motivated by the well-known Newton's second law) the inertial
 force $F_{inert}$ on a mass $m$ as seen by a gravitational
 source separated by a comoving distance $r_0$ is given by
 \be{eq1}
 F_{inert}\equiv mr_0\frac{\ddot{a}}{a}=mr_0\left(\dot{H}
 +H^2\right)=-mr_0\frac{4\pi G}{3}\left(\rho+3p\right)
 =mr_0\frac{4\pi G}{3}\left(2\rho+a\frac{d\rho}{da}\right),
 \ee
 in which we have used the energy conservation equation
 $\dot{\rho}+3H(\rho+p)=0$. A bound structure dissociates when
 the inertial force $F_{inert}$ (dominated by dark energy) is
 equal to the force $F_{bound}$ holding together this bound
 structure~\cite{r10,r11}. For the gravitationally bound
 structure whose mass is $M$, it dissociates when
 \be{eq2}
 F_{inert}=mr_0\frac{\ddot{a}}{a}=F_{bound}=G\frac{Mm}{r_0^2}
 =mr_0\omega_0^2\,,
 \ee
 namely
 \be{eq3}
 \frac{\ddot{a}}{a}=\omega_0^2=\frac{GM}{r_0^3}\,,
 \ee
 or equivalently
 \be{eq4}
 f(a)\equiv\frac{1}{\rho_0}\left(2\rho+a\frac{d\rho}{da}\right)
 =\frac{1}{\rho_0}\left(2\rho+\frac{d\rho}{d\ln a}\right)
 =\frac{2H_0^{-2}}{\Omega_0}\frac{GM}{r_0^3}\,,
 \ee
 where $\omega_0$ is the angular velocity; $\rho_0$ is the
 present density of dark energy;
 $\Omega_0\equiv (8\pi G\rho_0)/(3H_0^2)$ is the present
 fractional density of dark energy. Note that in this paper,
 the subscript ``0'' usually indicates the present value of
 corresponding quantity, and we have set $a_0=1$. In fact, one
 can check that the results Eqs.~(\ref{eq3}) and~(\ref{eq4})
 are coincident with the ones in the case of phantom dark
 energy whose EoS $w$ is a constant~\cite{r4} (we refer to
 e.g.~\cite{r14} for its detailed derivation; note that the
 result of~\cite{r4} is invalid for a non-constant $w$, while
 our results Eqs.~(\ref{eq3}) and~(\ref{eq4}) are still valid).

However, if the bound structure is massive enough to
 significantly affect the local spacetime metric, it is not
 accurate to express $F_{inert}$ in terms of FRW
 metric~\cite{r9,r10}. A more accurate method was presented
 in~\cite{r14}. Following~\cite{r14}, in the Newtonian limit,
 the interpolating metric of the local spacetime is given by
 \be{eq5}
 ds^2=\left[1-\frac{2GM}{a(t)\eta}\right]dt^2-a^2(t)\left[
 d\eta^2+\eta^2\left(d\theta^2+\sin^2\theta d\varphi^2\right)
 \right]\,,
 \ee
 where $\eta$ is the comoving radial coordinate. The radial
 equation of motion for a test particle in the Newtonian limit
 reads
 \be{eq6}
 \ddot{r}=\frac{\ddot{a}}{a}r+\frac{L^2}{r^3}-\frac{GM}{r^2}\,,
 \ee
 where $L=r^2\dot{\varphi}=r^2\omega=const.=r_0^2\omega_0$ is
 the constant angular momentum per unit mass. Noting that
 $\omega_0^2=GM/r_0^3\,$, Eq.~(\ref{eq6}) can be recast as
 \be{eq7}
 \ddot{r}=\frac{\ddot{a}}{a}r+\frac{GMr_0}{r^3}-\frac{GM}{r^2}
 =-\frac{dV_{\rm eff}}{dr}\,,
 \ee
 where the time-dependent effective potential is given by
 \be{eq8}
 V_{\rm eff}=-\frac{1}{2}\frac{\ddot{a}}{a}r^2+
 \frac{GMr_0}{2r^2}-\frac{GM}{r}\,.
 \ee
 The bound structure dissociates when the minimum of the
 time-dependent effective potential (including the centrifugal
 term) disappears. Note that the corresponding
 radius $r_{_{V_{min}}}$ at which the effective potential
 reaches its minimum is determined by
 \be{eq9}
 \frac{\ddot{a}}{a}r^4_{_{V_{min}}}-GMr_{_{V_{min}}}+GMr_0=0\,.
 \ee
 One can find (with the help of e.g. Mathematica) that
 Eq.~(\ref{eq9}) has a real solution only for
 \be{eq10}
 \frac{\ddot{a}}{a}\leq\frac{27}{256}\frac{GM}{r_0^3}\,.
 \ee
 Therefore, the minimum of the time-dependent effective
 potential disappears when
 \be{eq11}
 \frac{\ddot{a}}{a}=\frac{27}{256}\frac{GM}{r_0^3}\,,
 \ee
 which is also the time when the bound structure dissociates.
 Similarly, we can also recast Eq.~(\ref{eq11}) as
 \be{eq12}
 f(a)\equiv\frac{1}{\rho_0}\left(2\rho+a\frac{d\rho}{da}\right)
 =\frac{1}{\rho_0}\left(2\rho+\frac{d\rho}{d\ln a}\right)
 =\frac{27}{128}\frac{H_0^{-2}}{\Omega_0}\frac{GM}{r_0^3}\,.
 \ee
 As is shown in~\cite{r14}, the accurate disintegration time
 obtained here is usually earlier than the qualitative one
 obtained in~\cite{r4} (this point can also be seen by
 comparing Eqs.~(\ref{eq11}), (\ref{eq12}) with
 Eqs.~(\ref{eq3}), (\ref{eq4})). It is worth noting that most
 of the results in~\cite{r14} are valid for dark energy whose
 EoS $w$ is a constant, while our results obtained here are
 also valid for dynamical dark energy whose EoS $w$ is not a
 constant.

In the present work, we consider five bound structures, namely
 the Coma Cluster ($M=6\times 10^{48}\,{\rm g}$,
 $r_0=9\times 10^{24}\,{\rm cm}$), the Milky Way galaxy
 ($M=2\times 10^{45}\,{\rm g}$,
 $r_0=5\times 10^{22}\,{\rm cm}$), the Solar System
 ($M=2\times 10^{33}\,{\rm g}$,
 $r_0=7\times 10^{15}\,{\rm cm}$), the Earth
 ($M=6\times 10^{27}\,{\rm g}$,
 $r_0=6.4\times 10^8\,{\rm cm}$), and the Hydrogen atom. Note
 that the first three bound structures are suitable for
 Eqs.~(\ref{eq11}) and~(\ref{eq12}), while the Earth is
 suitable for Eqs.~(\ref{eq3}) and~(\ref{eq4}), since its
 surface material has no centrifugal term in the effective
 potential. Finally, the Hydrogen atom is also suitable for
 Eqs.~(\ref{eq3}) and~(\ref{eq4}), but in which the term $GM$
 should be replaced by a new term
 $q_e^2/(4\pi\varepsilon_0 m_e)$, since $F_{bound}$ is the
 electromagnetic force in this case. Note that
 $q_e=1.6\times 10^{-19}\,{\rm C}$, $m_e=0.511\,{\rm MeV}$
 ($1\,{\rm MeV}=1.7827\times 10^{-27}\,{\rm g}$),
 $1/(4\pi\varepsilon_0)=9\times 10^{18}\,{\rm g\,cm^3\,C^{-2}
 \,sec^{-2}}$, and $r_0=5.3\times 10^{-9}\,{\rm cm}$ for
 the Hydrogen atom. In Eqs.~(\ref{eq4}) and~(\ref{eq12}),
 $G=6.672\times 10^{-6}\,{\rm cm^3\,g^{-1}\,sec^{-2}}$,
 $H_0^{-1}=3.0856\times 10^{17}\,h^{-1}\,{\rm sec}$ (here $h$
 is the Hubble constant $H_0$ in units of
 $100\,{\rm km/sec/Mpc}$).  Note that $\Omega_0$ and $h$
 will be determined by the observational data (see below).

Since the current observational data are in the epoch $a<1$,
 we cannot ignore the contribution from pressureless matter
 in this stage, although it can be safely ignored in the above
 discussions when the universe is dominated by dark energy.
 From the Friedmann equation
 $3H^2=8\pi G\rho_{tot}=8\pi G(\rho+\rho_m)$ and
 $\rho_m=\rho_{m0}a^{-3}$, we have
 \be{eq13}
 E^2\equiv\frac{H^2}{H_0^2}=\Omega_0\frac{\rho}{\rho_0}+
 \Omega_{m0}a^{-3}=\Omega_0\frac{\rho}{\rho_0}
 +\left(1-\Omega_0\right)a^{-3}\,,
 \ee
 where $\Omega_{m0}\equiv(8\pi G\rho_{m0})/(3H_0^2)$ is the
 present fractional density of pressureless matter. If
 $\rho(a)$ is given, we can use Eq.~(\ref{eq13}) to constrain
 the model parameters with the current observational data. On
 the other hand, from Eqs.~(\ref{eq12}) or~(\ref{eq4}), one can
 find the corresponding scale factor $a_\ast$ at which the
 bound structure dissociates. Then, noting $H=\dot{a}/a$, we
 can evaluate the disintegration time measuring from today
 ($a=1$), namely
 \be{eq14}
 t_\ast-t_0=H_0^{-1}\int_1^{a_\ast}\frac{da}{aE(a)}=H_0^{-1}
 \int_0^{\ln a_\ast}\frac{d\ln a}{E(\ln a)}\,,
 \ee
 where $H_0^{-1}=9.7776\,h^{-1}\,{\rm Gyr}$.

%============================= section 3 ===================================

\section{An explicit model of Quasi-Rip}\label{sec3}

Here, we consider an explicit model of Quasi-Rip. In some
 sense, constructing a suitable model is a smart task. There
 are various ways to this end. For example, one can specify
 the scale factor $a$ as a function of the cosmic time $t$
 (see e.g.~\cite{r5,r15,r16}), the pressure $p$ as a function
 of the energy density $\rho$ (see e.g.~\cite{r6,r17,r18}),
 the energy density $\rho$ as a function of the scale factor
 $a$ (see e.g.~\cite{r9,r10}), or the Hubble parameter $H$ as
 a function of the cosmic time $t$ (see e.g.~\cite{r11}). It
 is shown in~\cite{r9,r10,r11} that these ways are equivalent
 in fact. In our case, for convenience, we choose to specify
 the energy density $\rho$ as a function of the scale factor
 $a$, similar to the cases of Little Rip~\cite{r9} and
 Pseudo-Rip~\cite{r10}. As mentioned in Sec.~\ref{sec1}, here
 our task is to construct an explicit function $\rho(a)$,
 which monotonically increases (namely $w<-1$) in the first
 stage and then monotonically decreases (namely $w>-1$)
 in the second stage. A naive idea is to use a piecewise
 function, which is phantom-like in the first sector (namely
 $\rho(a)\propto a^{-3(1+w_1)}$ with a constant EoS $w_1<-1$)
 and is quintessence-like in the second sector (namely
 $\rho(a)\propto a^{-3(1+w_2)}$ with a constant EoS $w_2>-1$).
 Then, let us refine this naive idea with a smooth function
 $\rho(a)\propto a^{\mu(a)}$, in which $\mu(a<a_t)>0$ and
 $\mu(a>a_t)<0$, where $a_t$ is the transition point. Noting
 that it is more convenient to use $\ln a$ as the variable
 in Eqs.~(\ref{eq12}), (\ref{eq4}), and~(\ref{eq14}), as well
 as the fact $a=e^{\,\ln a}$, it is appropriate to write $\mu$
 as a function of $\ln a$ instead. Therefore, the simplest
 function $\rho(a)$ can be given by
 \be{eq15}
 \rho(a)=\rho_0 a^{\alpha-\beta\ln a}
 =\rho_0 e^{\,\ln a\;(\alpha-\beta\ln a)}\,,
 \ee
 where $\alpha$ and $\beta$ are both constants. Let us have
 a closer observation. By requiring $d\rho/d\ln a=0$, we find
 that its extremum locates at $\ln a=\alpha/(2\beta)$. To
 ensure this extremum is a maximum, $d^2 \rho/d(\ln a)^2$
 should be negative here, therefore $\beta>0$ is required.

\begin{center}
 \begin{figure}[tbhp]
 \centering
 \includegraphics[width=0.49\textwidth]{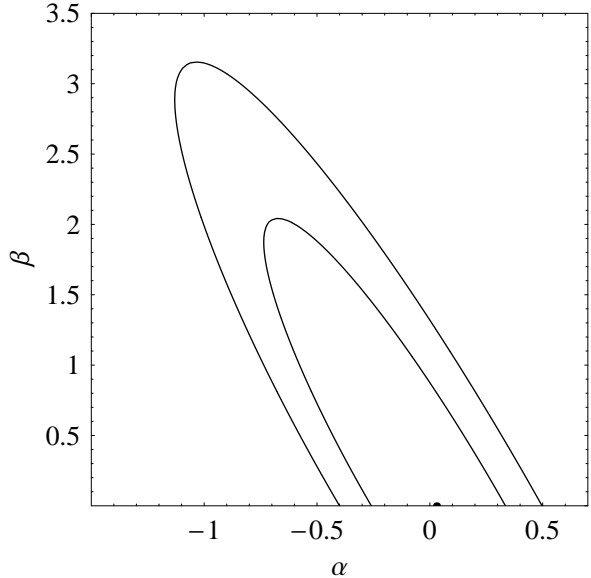}
 \caption{\label{fig1} The $68.3\%$ and $95.4\%$ confidence
 level contours in the $\alpha-\beta$ parameter space. The
 best-fit parameters are also indicated by a black solid
 point.}
 \end{figure}
 \end{center}

Substituting Eq.~(\ref{eq15}) into Eq.~(\ref{eq13}), we have
 \bea
 &E^2=\disp\frac{H^2}{H_0^2}&=\Omega_0\,a^{\alpha-\beta\ln a}
 +\left(1-\Omega_0\right)a^{-3}=\Omega_0\,e^{\,\ln a\;(\alpha
 -\beta\ln a)}+\left(1-\Omega_0\right)e^{-3\ln a}\nonumber\\
 & &=\Omega_0 (1+z)^{-\alpha-\beta\ln(1+z)}+
 \left(1-\Omega_0\right)(1+z)^3\,,\label{eq16}
 \eea
 where $z=1/a-1$ is the redshift. There are three free model
 parameters, namely $\Omega_0$, $\alpha$ and $\beta$. With
 this $E(z)$, following the methodology in
 e.g.~\cite{r19,r20,r21}, we can constrain this model with the
 latest Union2.1 Type~Ia supernovae (SNIa) dataset~\cite{r22}
 which consists of 580 SNIa. In fact, we find that the best fit
 has $\chi^2_{min}=562.224$, and the best-fit parameters are
 $\Omega_0=0.718859$, $\alpha=0.0294816$ and $\beta=0.0002525$,
 whereas the corresponding $h=0.698862$. In Fig.~\ref{fig1},
 we present the corresponding $68.3\%$ and $95.4\%$ confidence
 level~(C.L.) contours in the $\alpha-\beta$ parameter space.
 Although their best-fit values are small, $\alpha$~and $\beta$ can
 still be consistent with the observational data in the large
 $2\sigma$~C.L. region, i.e., $-1.2\,\lsim\,\alpha\,\lsim\,0.5$
 and $0\leq\beta\,\lsim\,3.2$.

%============================= Fig. 1 =================================

%======================================================================

\vspace{-4mm}  % used here just for a more comfortable typesetting

\begin{center}
 \begin{figure}[tbhp]
 \centering
 \includegraphics[width=0.98\textwidth]{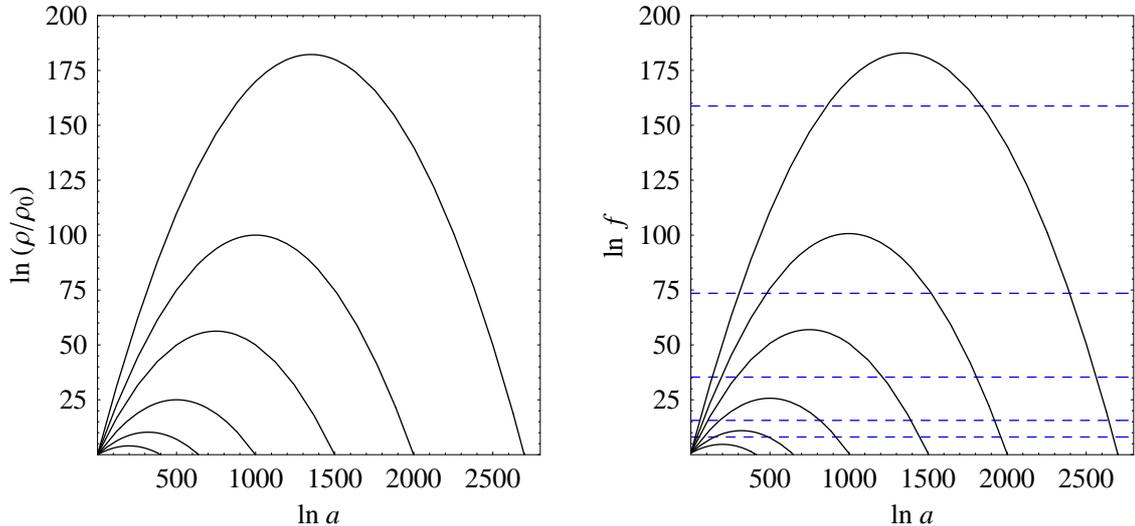}
 \caption{\label{fig2} $\ln\,(\rho/\rho_0)$ and $\ln f$ as
 functions of $\ln a$ for various model parameters. The curves
 from the innermost to the outermost are for $\alpha=0.04$,
 $\alpha=0.064$, $\alpha=0.1$, $\alpha=0.15$, $\alpha=0.2$
 and $\alpha=0.27$, respectively, while $\beta$ is fixed
 to be 0.0001. The dashed horizontal lines in the right
 panel indicate the corresponding thresholds to disintegrate
 the Coma Cluster, the Milky Way galaxy, the Solar System,
 the Earth and the Hydrogen atom (from bottom to top). Note
 that $\Omega_0$ and $h$ are taken to be their best-fit
 values.}
 \end{figure}
 \end{center}

Next, we show the Quasi-Rip in this model clearly. Substituting
 Eq.~(\ref{eq15}) into the reduced inertial force $f(a)$
 defined in Eqs.~(\ref{eq12}) and~(\ref{eq4}), we have
 \be{eq17}
 f(a)=\frac{\rho}{\rho_0}\left(2+\alpha-2\beta\ln a\right)
 =\left(2+\alpha-2\beta\ln a\right)a^{\alpha-\beta\ln a}
 =\left(2+\alpha-2\beta\ln a\right)
 e^{\,\ln a\;(\alpha-\beta\ln a)}\,.
 \ee
 In Fig.~\ref{fig2}, we plot $\ln\,(\rho/\rho_0)$ and $\ln f$
 as functions of $\ln a$ for various model parameters. For
 convenience, we fix $\beta=0.0001$ which is at the same order
 of its best-fit value, while the various values of $\alpha$
 are taken from the $1\sigma$~C.L. region in Fig.~\ref{fig1}.
 From the left panel of Fig.~\ref{fig2}, it is easy to see that
 our $\rho(a)$ in Eq.~(\ref{eq15}) is the desirable one, which
 monotonically increases in the first stage and
 then monotonically decreases in the second stage. It is not
 surprising that the plot of $\ln f$ is very similar to the
 one of $\ln\,(\rho/\rho_0)$, since their difference is given
 by $\ln\,(2+\alpha-2\beta\ln a)$ which is a small quantity in
 fact (cf. Eq.~(\ref{eq17})). Depending on the model parameters
 $\alpha$ and $\beta$, the inertial force can successively
 disintegrate the Coma Cluster, the Milky Way galaxy, the Solar
 System, the Earth and the Hydrogen atom. In different cases,
 some bound structures will be disintegrated while the other
 bound structures will not. If none of the bound structures is
 disintegrated, the corresponding Quasi-Rip is a failed rip.
 From the right panel of Fig.~\ref{fig2}, we find that the
 innermost one is a failed rip, the outermost one is the
 strongest Quasi-Rip which can disintegrate all the five bound
 structures, and the others are moderate ones which can only
 disintegrate one or several of the five bound structures.
 From Eqs.~(\ref{eq12}) or~(\ref{eq4}), we can find the
 corresponding $\ln a_\ast$ at which the bound structure
 dissociates. Then, we can evaluate the disintegration time
 measuring from today $t_\ast-t_0$ by using Eq.~(\ref{eq14})
 with $E(\ln a)$ given in Eq.~(\ref{eq16}), and present the
 results in Tab.~\ref{tab1}. For a fixed $\beta$, a bound
 structure dissociates earlier with a larger $\alpha$, since
 the corresponding inertial force is stronger. In the case of
 $\alpha=0.27$, the difference between the disintegration time
 of Earth and Solar System is 4349 years, while the difference
 between the disintegration time of Hydrogen atom and Earth is
 only 15 years. Finally, it is easy to see that the most
 distinct feature of Quasi-Rip is that the inertial force
 monotonically decreases in the second stage. Eventually, it
 will become lower than all the thresholds to disintegrate
 the bound structure. Therefore, the already disintegrated
 structures have the possibility to be recombined in the second
 stage. This is the unique feature of Quasi-Rip different from
 Big Rip, Little Rip and Pseudo-Rip. Our universe has a chance
 to be rebuilt from the ashes after the terrible rip. This
 might be the last hope in the ``hopeless'' rip.

%============================= Fig. 2 =================================

%======================================================================

\vspace{-7mm}  % used here just for a more comfortable typesetting

%=================== table 1 ===================

 \begin{table}[tbhp]
 \begin{center}
 \begin{tabular}{l|llllll}
 \hline\hline\\[-4.1mm]
 Bound structure~~ & ~~$\alpha=0.04$~~ & $\alpha=0.064$~~
 & $\alpha=0.1$~~~ & $\alpha=0.15$~~ & $\alpha=0.2$
 & $\alpha=0.27$ \\[0.5mm]
 \hline\\[-4mm]
 Coma Cluster & & 563.154 & 333.555 & 216.991
  & 161.28 & 118.715\\[1mm]
 Milky Way & & & 344.035 & 223.126 & 165.75 & 122.005\\[1mm]
 Solar System & & & & 223.281 & 165.857985136 & 122.083461362\\[1mm]
 Earth & & & & & 165.857991583~~~ & 122.083465711\\[1mm]
 Hydrogen atom & & & & & & 122.083465696~\\[1mm]
 \hline\hline
 \end{tabular}
 \end{center}
 \caption{\label{tab1} The disintegration time measuring from
 today $t_\ast-t_0$ (in units of Gyr) for models with various
 $\alpha$, while $\beta$ is fixed to be 0.0001, and $\Omega_0$
 and $h$ are taken to be their best-fit values.}
 \end{table}

%===============================================

\vspace{-4mm}  % used here just for a more comfortable typesetting

%============================= section 4 ===================================

\section{Concluding remarks}\label{sec4}

The fate of our universe is an unceasing topic of cosmology
 and the human being. The discovery of the current accelerated
 expansion of the universe significantly changed our view of
 the fate of the universe. Recently, some interesting scenarios
 concerning the fate of the universe attracted much attention
 in the community, namely the so-called ``Little Rip'' and
 ``Pseudo-Rip''. It is worth noting that all the Big Rip,
 Little Rip and Pseudo-Rip arise from the assumption that the
 dark energy density $\rho(a)$ is monotonically increasing. In
 the present work, we are interested to investigate what will
 happen if this assumption is broken, and then propose a
 so-called ``Quasi-Rip'' scenario, which is driven by a type
 of quintom dark energy. In this work, we consider an explicit
 model of Quasi-Rip in detail. We show that Quasi-Rip has an
 unique feature different from Big Rip, Little Rip and
 Pseudo-Rip. Our universe has a chance to be rebuilt from the
 ashes after the terrible rip. This might be the last hope in
 the ``hopeless'' rip.

Some remarks are in order. Firstly, as is shown in
 Sec.~\ref{sec3}, our Quasi-Rip model is well consistent with
 the current observational data. However, even in the $1\sigma$
 C.L. region of $\alpha-\beta$ parameter space, the future behavior
 of our universe can be different enough (depending on the
 particular model parameters $\alpha$ and~$\beta$). In fact,
 as is well known, the current observational data can be
 consistent with all the phantom-like, quintessence-like and
 quintom-like dark energy models. Therefore, the current
 observational data cannot tightly tell what is the true fate
 of our universe. Most of the possibilities (including Big
 Rip, Little Rip, Pseudo-Rip, Quasi-Rip, de~Sitter expansion,
 other future singularities and so on) are still living.

Secondly, the explicit model of Quasi-Rip considered in the
 present work is the simplest case. One can construct other
 more complicated $\rho(a)$ to implement the Quasi-Rip. For
 example, one might construct an EoS $w(a)$ as a function
 of scale factor $a$, which is smaller than $-1$ when $a<a_t$
 and larger than $-1$ when $a>a_t$. Then the corresponding
 $\rho(a)$ can be found from the energy conservation equation
 $\dot{\rho}+3H\rho(1+w(a))=0$. Of course, other smart methods
 to construct the desirable $\rho(a)$ are awaiting us.

Thirdly, as mentioned in~\cite{r10}, in its second Pseudo-Rip
 model, the reduced inertial force $f(a)$ can also have a peak,
 similar to our Quasi-Rip model. However, we note that in the
 Pseudo-Rip model, after the peak, the reduced inertial force
 $f(a)\to const.$ which is still higher than the corresponding
 threshold to disintegrate the bound structure. Therefore, the
 already disintegrated structures have {\em no} possibility
 to be recombined in the Pseudo-Rip models. On the contrary, in
 the Quasi-Rip models, the reduced inertial force $f(a)$
 monotonically decreases in the second stage. Eventually, it
 will become lower than all the thresholds to disintegrate the bound
 structure. Therefore, the already disintegrated structures
 have the possibility to be recombined in the second stage.

Fourthly, it is well known that phantom is unstable at quantum
 level and hence the perturbations grow large. Noting that
 in the present work our discussions are at classical level
 instead, this problem could be set aside. In fact, the quantum
 stability of a phantom phase has been considered in~\cite{r24}. The
 authors of~\cite{r24} studied the perturbations in the
 quantum-corrected effective field equation at one- and
 two-loop order, and they found that the system is stable. On
 the other hand, it is claimed in~\cite{r25} that scalar
 perturbations can grow during a phantom phase if EoS $w<-5/3$.
 However, from Eq.~(\ref{eq15}) and Fig.~\ref{fig1}, it is easy
 to see that the corresponding $w$ is only slightly smaller
 than $-1$ in the phantom phase for our particular $\alpha$ and
 $\beta$ (n.b. Fig.~\ref{fig1}). So, $w>-5/3$ instead and hence our
 Quasi-Rip model can avoid the problem raised in~\cite{r25}.
 Further, in fact the dark energy considered in this work is
 not necessarily a scalar field. It could even be an
 effective dark energy from modified gravity, namely the
 so-called ``geometric dark energy''. So, it might avoid the
 corresponding problems in the phantom phase.

Finally, in the present work, we consider only the quintom
 dark energy which crosses the phantom divide $w=-1$ once.
 In fact, we can also consider the quintom dark energy which
 can cross the phantom divide for many times. The most
 attractive Quasi-Rip model might be the one driven by the
 oscillatory quintom dark energy. In this oscillatory Quasi-Rip
 model, our universe will be destroyed and then be rebuilt
 again and again.

%============================= acknowledgements ===================================

\section*{ACKNOWLEDGEMENTS}
We thank the anonymous referee for quite useful comments and
 suggestions, which helped us to improve this work. We are
 grateful to Professors Rong-Gen~Cai and Shuang~Nan~Zhang
 for helpful discussions. We also thank Minzi~Feng for kind
 help and discussions. This work was supported in part by NSFC
 under Grants No.~11175016 and No.~10905005, as well as NCET
 under Grant No.~NCET-11-0790, and the Fundamental Research
 Fund of Beijing Institute of Technology.

\renewcommand{\baselinestretch}{1.1}

%============================= references ==================================

\end{document}